# On the State of Water Ice on Saturn's Moon Titan and Implications to Icy Bodies in the Outer Solar System

Weijun Zheng,[1, 2, 3 *] David Jewitt,[1 *] and Ralf I. Kaiser[2 *]

[1] *Institute for Astronomy, University of Hawaii, Honolulu, HI 96822*

[2] *Department of Chemistry, University of Hawaii, Honolulu, HI 96822*

[3] present address*: Beijing National Laboratory for Molecular Sciences, State Key Laboratory of Molecular Reaction Dynamics, Institute of Chemistry, Chinese Academy of Sciences, Beijing 100190, P. R. China*

[*] Authors to whom the correspondence should be sent:

ralfk@hawaii.edu, jewitt@hawaii.edu, zhengwj@iccas.ac.cn

**Abstract**

The crystalline state of water ice in the Solar System depends on the temperature history of the ice and the influence of energetic particles to which it has been exposed. We measured the infrared absorption spectra of amorphous and crystalline water ice in the 10-50 K and 10-140 K temperature range, respectively, and conducted a systematic experimental study to investigate the amorphization of crystalline water ice via ionizing radiation irradiation at doses of up to 160 ± 30 eV per molecule. We found that crystalline water ice can be converted only *partially* to amorphous ice by electron irradiation. The experiments showed that a fraction of the 1.65 μm band, which is characteristic for crystalline water ice, survived the irradiation, to a degree that *strongly* depends on the temperature. Quantitative kinetic fits of the temporal evolution of the 1.65 μm band clearly demonstrate that there is a balance between thermal recrystallization and irradiation-induced amorphization, with thermal recrystallizaton dominant at higher temperatures. Our experiments show the amorphization at 40K was incomplete, in contradiction to Mastrapa and Brown's conclusion (*Icarus* **2006**, *183*, 207.). At 50 K, the recrystallization due to thermal effects is strong, and most of the crystalline ice survived. Temperatures of most icy objects in the Solar System, including Jovian satellites, Saturnian satellites (including Titan), and Kuiper Belt Objects, are equal to or above 50 K; this explains why water ice detected on those objects is mostly crystalline.

## 1. Introduction

The atmosphere of Saturn's moon Titan holds a rich chemical inventory ranging from its main constituents, i.e. molecular nitrogen ($N_2$) and methane ($CH_4$), to hydrocarbon molecules such as acetylene ($C_2H_2$), ethylene ($C_2H_4$), ethane ($C_2H_6$), methylacetylene ($CH_3CCH$), propane ($C_3H_8$), diacetylene ($C_4H_2$), and benzene ($C_6H_6$),[1,2] nitriles involving hydrogen cyanide (HCN), cyano-acetylene (HCCCN), and cyanogen ($C_2N_2$),[1,2] as well as oxygen-bearing molecules carbon dioxide ($CO_2$), carbon monoxide (CO), and water ($H_2O$). In the solid state, dicyanoacetylene (NCCCCN, $C_4N_2$) was detected on Titan via its $\nu_8$ fundamental at 478 $cm^{-1}$.[3] One decade later, Khanna et al. [4] suggested the presence of solid cyanoacetylene (HCCCN). Coustenis et al. [5] and Samuelson et al. [3] indicated that also ethane ($C_2H_6$), acetylene ($C_2H_2$), cyanoacetylene (HCCCN), and possibly hydrogen cyanide (HCN) could exist in the solid state. Very recently, a detailed analysis of the 6711 $cm^{-1}$ (1.49 μm), 5000 $cm^{-1}$ (2.0 μm), and 3448 $cm^{-1}$ (2.9 μm) infrared absorption bands strongly indicates that solid water ice also exists on Titan.[6,7] However, it has not been determined whether the water ice on Titan is in crystalline or amorphous form. In this context, it is important to note that the surface temperature on Titan might have exceed 300 K during the first ~$10^8$ years of its history.[8] Therefore, 'primordial' water ice on Titan might be crystalline unless amorphous ice can be produced by interaction of high energy galactic cosmic ray particles penetrating through Titan's atmosphere to the surface.

In this context, it is important to stress that amorphous water ice converts spontaneously to crystalline ice on a timescale that depends exponentially on the temperature via $\tau_{cr} = 3.0 \times 10^{-21} e^{E_A/kT}$. Here, $\tau_{cr}$ is the crystallization time in years at temperature $T$ [K], $k$ is Boltzmann's constant and $E_A$ is the activation energy; $E_A/k$ equals 5370 K.[9] In the warm inner solar system, including the familiar terrestrial environment of the Earth, the crystallization is nearly instantaneous and all natural ice is crystalline. However, further out in the solar system, the crystallization time lengthens dramatically. Water ice on the three outermost Galilean satellites of Jupiter (at about 5 AU the surface temperatures are between 80 K and 150 K [10,11]) is partly amorphous.[12] The spatial distribution of the amorphous ice is correlated with the flux of energetic particles trapped within Jupiter's extensive magnetosphere, suggesting that amorphous ice is produced on these satellites via magnetospheric interaction. Both amorphous and crystalline water ice exist on Saturn's satellite Enceladus.[13,14] Also, very recently, water ice on Kuiper Belt Objects (KBOs), such as Quaoar and 2003 $EL_{61}$, has been found to be crystalline.[15-17] The prime observational diagnostic is the 6060 $cm^{-1}$ (1.65) μm band that is a feature of crystalline (but not amorphous) water ice. The surface temperature on Quaoar is close to 50 K [15] and, by the above equation, water ice on Quaoar should be indefinitely stable in the amorphous phase. The unexpected presence of crystalline ice leads to two conclusions.[15] First, the ice must have been formed and/or heated to temperatures higher than about 100 K, substantially above the radiation equilibrium temperature. Second, this heating must be recent, because the surface of Quaoar is exposed to energetic particles from the solar wind and the cosmic rays that should amorphize the exposed ice on a timescale much shorter than the age of the solar system. Based on published experimental data, Jewitt and Luu [15] estimated an age 1 to 10 Myr for the surface ice. Subsequent experiments by Mastrapa and Brown [18] support a 1 Myr timescale. The source of the heating remains unidentified: suggested possibilities range from heating by the impact of micrometeoroids to ejection of previously buried, warm ice by on-going cryovolcanic activity.

Therefore, the actual state of water ice, (i.e. crystalline versus amorphous) is of critical importance to understand the formation temperature and exposure of water ices to ionizing radiation not only to Saturn's moon Titan but in the outer Solar System in general. Due to this importance, the amorphization of crystalline water ice has been investigated by many groups. Kouchi and Kuroda [19] found that cubic ice below 70 K can be transformed into amorphous ice by UV photons. Moore and Hudson [20] found that crystalline ice can be converted to an amorphous phase when irradiated by 700 keV protons at temperatures between 77 and 13 K. They also indicated that the conversion rate increases as the temperature is decreased and the converted fraction is dose dependent. Strazzulla et al. [21] studied the transition from crystalline to amorphous water induced by keV ion irradiation at temperatures between 10 and 100 K. Leto and Baratta [22] examined the UV Lyman-α photon amorphization of crystalline water ice based on the 3.0 μm band. Very recently, Mastrapa and Brown [18] investigated the effect of irradiation on the 1.65 μm band of water ice. As already noted, this band is a very important feature for distinguishing crystalline ice from amorphous ice.[23] However, more systematic laboratory studies are still needed to investigate the amorphization of crystalline ice by irradiation in order to provide detailed mechanistical information and the temperature, dose, and particle-dependent amorphization rates needed to understand the results from astronomical observations. In this work, we first compare the near infrared spectra of cubic crystalline ice and amorphous ice and investigate thereafter the effects of electron irradiation on the near infrared spectra of crystalline water ice.

## 2. Experimental

The experiments were carried out in an ultrahigh vacuum chamber ($5\times10^{-11}$ torr).[24] Briefly, a two-stage closed-cycle helium refrigerator coupled with a rotary platform is attached to the main chamber and holds a polished polycrystalline silver mirror serving as a substrate for the ice condensation. With the combination of the closed-cycle helium refrigerator and a programmable temperature controller, the temperature of the silver mirror can be regulated precisely (± 0.3 K) between 10 K and 350 K. A valve and a glass capillary array are used to condense gases on the silver mirror. The actual thickness of the ice samples can be controlled via the condensation time and the water partial pressure in the main chamber. The cubic crystalline water ice was prepared by condensing water vapor onto the silver substrate at 140 K for 7 minutes and cooling to the desired temperatures (10 – 50 K) with a cooling rate of 1.0 K $min^{-1}$. The amorphous water ice was prepared by depositing water onto the silver mirror at 10 K for 5 minutes, then warming it up to the desired temperature at 0.5 K $min^{-1}$. The thickness of the samples was determined to be 370 ± 50 nm. It is also important to comment on the determination of the thicknesses of the deposited samples. Typically, this is done by utilizing a modified Lambert-Beer law.[25] However, our measurements were done in the absorption-reflection-absorption mode, but integrated absorption coefficients to determine the absolute thickness of the samples were mainly measured in transmission.[26] Here, the profiles of the infrared bands in transmission might be different from those recorded in the absorption-reflection-absorption mode. This also could translate into different absorption coefficients, where some absorption coefficients are affected more than others. To bypass this problem, we utilized our mass spectral data of the non-irradiated sample to determine the thickness of the sample.[23,27] Briefly, in this blank experiment, the condensed ice sample was warmed up, and ion profiles were recorded at m/z = 18 ($H_2O^+$); note the water molecule can also undergo dissociative ionization by the electrons of the electron impact ionizer to yield signal at, for instance, m/z = 17 ($HO^+$) or m/z = 16 ($O^+$). Separate experiments with pure water samples showed that about 25 ± 4 % of the water molecules undergo dissociative ionization. Integrating the signal at m/z = 18, correcting for the fragmentation patterns, we yield integrated ion signals of $2.2\times10^{-4}$ As. Taking into account the pumping speed of our system, the sample thicknesses derive to be 370 ± 50 nm. These crystalline ice samples were irradiated with 5 keV electrons for 44 hours at beam currents of 200 nA. The infrared spectra of the samples were measured *on-line* and *in-situ* by a Fourier transform infrared spectrometer (Nicolet 6700 FTIR) at 10000-2000cm$^{-1}$ (1-5 μm) range with the absorption–reflection–absorption method. Each spectrum was averaged for 6 minutes.

## 3. Results

### 3.1. Infrared spectra of crystalline and amorphous ice

Figure 1 shows the infrared reflection-absorption spectra of cubic crystalline water ice between 10 and 140 K. For convenience, we name the absorption between 2000 and 2600cm$^{-1}$ "4.5 μm band", the absorption between 2800 and 3800cm$^{-1}$ "3 μm band", the absorption between 4400 and 5400 cm$^{-1}$ "2.0 μm band", and the absorption between 5600 and 7400 cm$^{-1}$ "1.5 μm band". The most striking changes are the temperature dependences of the 3 μm band and the 1.5 μm band. At 10 K, the 3 μm band has at least four major components centered at 3096, 3166, 3330, and 3457 cm$^{-1}$. It is broadened with increasing temperature and its components are obscured at higher temperatures. The 1.5 μm band at 10 K has at least three major components centered at 6028, 6327, and 6651 cm$^{-1}$. These components are relatively obscure at higher temperatures. The intensity of the 1.65 μm band (6060 cm$^{-1}$) decreases dramatically as the temperature increases from 10 K to 140 K, in agreement with Grundy et al..[28] For comparison, Figure 2 displays the infrared absorption spectra of amorphous water ice between 10 and 50 K. The absorption features at 3718 and 3693 cm$^{-1}$ correspond to the dangling OH bonds of the 2- and 3-coordinated water molecules in the micro-pores of the amorphous ice.[29,30] The absorption feature at 5313 cm$^{-1}$ has not been reported before. Here, we assign it to the stretching of the dangling OH bond coupled with the bending of the water molecules.[31] The existence of the dangling OH bonds suggests that the water ice deposited at 10 K was porous amorphous ice. We warmed up the amorphous ice from 10 K to 50 K at a speed of 0.5 K min$^{-1}$. The intensities of the dangling bond absorptions decrease with increasing temperature, implying that the annealing of the amorphous ice caused the collapse of the micro-pores and, therefore, reduced the number of OH dangling bonds. We also notice that the center of the 3 μm band decreases with increasing temperature. Note that the 4.15 μm peak originated from the O-D stretch mode of HDO present in natural abundance in the ice; it is not prominent in amorphous ice because of its large bandwidth. The temperature shift of HDO 4.15 μm peak has also been reported by Rice and co-workers.[32] The temperature dependence of the infrared absorption observed in our experiments is consistent with previous studies.[18,33]

We summarize now the major empirical differences between the infrared spectra of crystalline and amorphous ice. First, crystalline ice has features at 1.65 μm (6060 $cm^{-1}$), 2.35 μm (4248 $cm^{-1}$), and 4.15 μm (2411$cm^{-1}$) that are absent in the spectra of amorphous ice. Second, the 3 μm band of the crystalline ice has more fine structure than that of amorphous ice. Third, the (porous) amorphous ice holds dangling bond features at 2.708 μm (3693 $cm^{-1}$), 2.690 μm (3718 $cm^{-1}$), and 1.882 μm (5313 $cm^{-1}$), which are absent in the spectra of crystalline ice. Fourth, the temperature shifts of the infrared features are opposite between crystalline ice and amorphous ice. The most significant difference with respect to astronomical observations the 1.65 μm band, since this band falls in a relatively clean portion of the terrestrial atmospheric transmission spectrum and can be readily observed. We will pay special attention to that band in the irradiation experiments.

### 3.2. The Electron irradiation of cubic crystalline ice

We studied the irradiation of cubic crystalline water ice at temperatures of 10, 20, 30, 40, and 50 K. Figure 3 shows the infrared reflection-absorption spectra of cubic crystalline water ice before and after irradiation. The irradiation at 10 K changed the absorption of crystalline ice dramatically. The maximum of the 1.5 μm band blue shifted from 6651 to 6705 $cm^{-1}$, the second component of the 1.5 μm band at 6327 $cm^{-1}$ is gone, and the 1.65 μm band almost disappeared after irradiation. The irradiation broadened the 3 μm band and destroyed its fine structure. Irradiation also caused other changes, i.e. the 2.0 μm band blue shifted from 4900 to 4980 $cm^{-1}$, and the maximum of the 4.5 μm band red shifted from 2248 to 2200 $cm^{-1}$. The irradiation also caused the disappearance of the 2.35 μm (4248 $cm^{-1}$) and of the 2411 $cm^{-1}$ feature. Overall, the spectra after irradiation are very similar to the spectra of amorphous ice except lacking in OH dangling bond features. This indicates that the cubic crystalline ice has been converted to amorphous ice in the 10 K experiment, but the absence of the OH dangling bonds tells us that the amorphous ice resulting from irradiation is non-porous amorphous ice. Note that the spectra after irradiation show an additional absorption band centered at 2851 $cm^{-1}$. This feature is from hydrogen peroxide ($H_2O_2$) generated by the irradiation.[25] Moore et al. [17] reported an oscillation between crystalline and amorphous phases during irradiation, a phenomenon that has not been apparent in our experiments.

Next, we compare the experiments at different temperatures (10-50K). We can see from Figure 3 that the shifts of the 1.5 μm, 2.0 μm and 4.5 μm bands induced by irradiation become less significant at higher temperatures. The second component of the 1.5 μm band at 6327 $cm^{-1}$ disappeared in the 20 and 30 K experiments, but it survived the irradiation at 40 and 50 K although its intensity has been decreased by the irradiation. Also, the change of the 2.35 μm feature is not obvious at higher temperatures, such as at 40 and 50 K. The most important differences are in the 1.65 μm band and the 3 μm band. Apparently, more of the 1.65 μm band survived the irradiation at the higher temperatures. The irradiation destroyed the fine structure of the 3 μm band at 10 K, but caused only very little change to the same band when the temperature was increased to 50 K. After the irradiation, we also warmed up and to recooled the ice samples. The irradiated samples were warmed up to 145 K at 0.5 K $min^{-1}$ and kept at 145 K for 1 hour for recrystallization (at this temperature, the above mentioned equation gives $\tau_{cr}$ = 0.32 hours; therefore, the crystallization is complete). Then, the ices were cooled down back to the original temperature. Figure 4 shows a comparison of the spectra before irradiation, after irradiation, and after recooling. The spectra from the recooled samples are almost identical to those before irradiation except that the recooled samples show a hydrogen peroxide feature that resulted from the irradiation.

## 4. Discussion

The survival of the 1.65 μm band after irradiation indicates that the amorphization of the crystalline ice is not complete, especially as the temperature exceeds 20 to 30 K (Figure 3). In order to quantify the results, we measured the area of the 1.65 μm band during and after irradiation, then compared it to the area before irradiation. Here, we define the ratio between the surviving and original 1.65 μm band areas as α: $\alpha = \sigma_s/\sigma_o$, where, $\sigma_s$ is the area of the surviving 1.65 μm band, and $\sigma_o$ is the area of the 1.65 μm band before irradiation. The band areas were measured by drawing the baseline along the slope underneath the 1.65 μm band, in the same way as done by previous researchers.[23,34,35] Figure 5 shows α versus irradiation time, for ice in the 10 K to 50 K temperature range. Figure 5 also presents the corresponding Kuiper Belt Object exposure time, which will be discussed in the following paragraphs. The 1.65 μm band decreases rapidly in strength at the beginning of the irradiation, then reaches an equilibrium that is related to the temperature. The change of α versus irradiation time, *t,* can be fitted with equation (1). The constants $A$ and $k_1$ for different temperatures are summarized in Table 1.

$$\alpha(t) = 1 - A \times (1 - e^{-t/k_1}) \qquad (1)$$

The band strength variations reveal that there is a balance between the thermal recrystallization and irradiation amorphization and the result of the balance clearly shifts with the temperature. A higher fraction of the 1.65 μm band survived the irradiation at the higher temperatures. Strazzulla et al. [21] performed a similar experiment with an ion beam (3 keV $He^+$) amorphization of crystalline water ice at different temperatures based on the 3-μm band. Our result on the 1.65-μm band is in agreement with the temperature-dependence of amorphization found in their experiments. Mastrapa and Brown[18] found that irradiation of crystalline ice with 0.8MeV protons produced the amorphous ice spectrum at low temperatures (9, 25, and 40 K). However, at 50 K, some crystalline absorptions persisted after irradiation. Moreover, the crystalline spectrum showed only slight changes after irradiation at 70 and 100 K. Our experiment is in agreement with theirs at high temperatures (> 50 K), but our low temperature results are in contradiction. Whereas they reported complete amorphization of crystalline ice at 9, 25 and 40 K, our experiment shows that some of the 1.65 μm band survived the irradiation in the 10–50 K temperature range. Therefore, we cannot conclude that the amorphization of crystalline ice was complete at 10–40K.

Table 1. The constants *A* and *k* fitted in Figure 5.

| Temperature | 10 K | 20 K | 30 K | 40 K | 50 K |
|---|---|---|---|---|---|
| $A$ | 0.91 ±0.06 | 0.81 ±0.06 | 0.79 ±0.06 | 0.65 ±0.06 | 0.39 ±0.06 |
| $k_1$ (h) | 3.0 ±0.8 | 2.3 ±0.5 | 3.0 ±0.5 | 3.5 ±0.5 | 2.4 ±0.5 |

In order to solve the contradiction, we carefully compared our spectra to Mastrapa and Brown's spectra. Since they did not report their spectra from the 9 and 25 K irradiation, here we focus on the spectra at 40 and 50 K. Our crystalline ice spectra at 40 and 50 K are very similar to theirs at the same temperatures. Our post-irradiation spectra are also similar to their post-irradiation spectra (all show the 1.65 μm feature). The 1.65 μm feature did persist in their irradiation experiment as well as in ours. The main difference lies between their spectra of amorphous ice and ours. We could not perceive any 1.65-μm feature in our amorphous ice spectra. Yet, in their amorphous ice spectra, at 40 and 50 K, the 1.65 μm feature is quite obvious. We propose that their amorphous ice had been contaminated by crystalline ice. Also, the 1.65 μm feature in their 40 K amorphous spectrum is even stronger than at 50 K. This runs against the trend observed in the current work.

The flux of the electrons in our experiments was about $4.0\times10^{11}$ electrons $cm^{-2}$ $s^{-1}$ while the density of crystalline water ice is $0.93 \pm 0.02$ g $cm^{-3}$ [36]. The LETs (Linear Energy Transfer) of 5 keV electrons in 370 nm thick ice simulated with CASINO [37] is 7.88 keV $\mu m^{-1}$. From these values, we compute that the dose in 44 hours is 160 eV per molecule. In our experiments, the amorphization of water ice reached equilibrium in a few hours (6-30 eV per molecule). Strazzulla et al. [21] found that 3 keV $He^+$ amorphization of water ice reached an equilibrium at a dose of ~10 eV per molecule. Mastrapa and Brown [18] found the dose is between 11 and 16 eV per molecule with 0.8 Mev protons. In addition, Leto and Baratta [22] discovered that the UV Lyman-α induced amorphization rate of water ice is about the same as those of fast ions ($H^+$, $He^+$, and $Ar^{++}$) at similar doses. Their results also show that the amorphization reached equilibrium at a dose of few eV per molecule. Overall, the laboratory results with different particles are comparable. Table 2 summarizes the previous experiments on infrared spectroscopic studies of amorphization of crystalline water by high energy particle irradiation. Table 3 shows the LETs of these particles in water ice calculated with SRIM [38,39] and CASINO [37]. The results in Table 3 indicate that the LETs of different particles in water ice can vary by a few orders of magnitudes. However, the irradiation experiments conducted with these particles all show that amorphization of water ice reaches equilibrium at a narrow dose range. This may imply the amorphization of crystalline water ice is mainly dose-dependent, which is reasonable since the main differences between distinct particles have already been considered by theory during the calculation of doses.

On the other hand, most energy transfer and destruction in water ice during irradiation occurs through secondary processes by interactions with secondary electrons, ions, or fragments. When water ice is irradiated with different particles, although the very first step of the water-particle-interaction is particle-dependent, the secondary processes are rather similar.

Recall that the temperatures of Jovian Satellites have been estimated to be 80 – 150 K [10,11] and those of Saturn's satellites to be 70 – 110 K [40,41]. This range also covers the surface temperature of Titan of 94 K. These temperatures are much higher than used in our experiments. Given that thermal recrystallization already occurs at 50 K, we anticipate that thermal crystallization will overcome some of the effects of irradiation and that the 1.65 μm band should be always present in the spectra of the Jovian and Saturnian satellites. Indeed, the spectra of Jupiter's and Saturn's satellites measured by ground and space-based telescopes and flyby spacecraft all show the 1.65 μm band. Hansen and McCord [12] found that the water ice on the three outermost Galilean satellites has a lattice structure that can vary with location from crystalline to amorphous. The amorphous form is produced by particle radiation in the magnetosphere. Model comparisons show that the ice on Europa is dominated by radiative disruption, that on Callisto is dominated by thermal crystallization, that of Ganymede is in between. Concerning Saturn's satellites, it has been suggested that amorphous ice might exist near Enceladus' poles, where the particle flux is high and the temperature is relatively low [13,14]. As a matter of fact, using 1 eV to GeV protons in their model, Cooper et al. [42] estimated it would take about $10^8$ years for the optical thickness (0.1 mm) of the ice on the surface of a Kuiper belt object to reach a dose of ~10 eV per molecule. Therefore, the dose from 44 hours irradiation in our experiment corresponds roughly to 1.6 billion years on the surface of a Kuiper belt object. We can see the 1.65-μm band of crystalline ice will survive more than 200 hours irradiation at 50K. That corresponds to about 6.3 billion years on the surface of a Kuiper belt object, longer than the history of our solar system. Once crystallized, ice on Quaoar and the other KBOs at similar temperatures will never be completely amorphized. Of course, the conditions on Quaoar and other objects are much more complex than our laboratory simulations, and we cannot rule out active crystallization processes, such as impact or cryovolcanic outgassing. We only conclude based on our experiments that these other processes are not needed to explain the observed persistence and prevalence of crystalline ice in the middle and outer regions of the solar system. Considering Titan, near its surface, the energy deposition from cosmic rays is about $3.16\times10^{-10}$ erg $cm^{-3}$ $s^{-1}$,[43] giving a dose of 1.68 eV $molecule^{-1}$ in 4 billion years. This dose equals to only 0.5 hour irradiation in our experiments, which means most of the water ice on Titan's surface should be crystalline.

Table 2. Previous infrared spectroscopy studies of amorphization of crystalline water by irradiation.

| Reference | Particle | Energy | Thickness (μm) | Temperature (K) | Dose (eV $molecule^{-1}$) | Band Wavelength |
|---|---|---|---|---|---|---|
| [44] | $He^+$ | 3 keV | ~0.01 | 10, 77 | 0 – 50 | 3 μm |
| [21] | $He^+$ | 3 keV | ~0.01 | 10, 20, 30, 55, 77, 100 | 2 – 10 | 3 μm |
| [20] | $H^+$ | 700 keV | ~4 | 13, 36, 46, 56, 66, 77 | 0 – 30 | 45 μm |
| [22] | UV Ly-α | 10.2 eV | ~0.05 | 16 | 0 – 100 | 3 μm |
| | $H^+$ | 30 keV | | | | |
| | $He^+$ | 30 keV | | | | |
| | $Ar^{++}$ | 60 keV | | | | |
| [18] | $H^+$ | 800 keV | 8 ± 1.5 | 9, 25, 40, 50, 70, 100 | 0.6 – 16 | 1.65 μm |
| This work | $e^-$ | 5 keV | 0.37 ± 0.05 | 10, 20, 30, 40, 50 | 0 – 160 | 1.65 μm |

Table 3. Nuclear and inelastic stopping powers of various particles into 100 nm thick water ice.

| Irradiating particle | LET($S_e$), keV $\mu m^{-1}$ | LET($S_n$), keV $\mu m^{-1}$ |
|---|---|---|
| $e^-$, 5 keV | 5.52 | – |
| $H^+$, 30 keV | 65.2 | 0.36 |
| $H^+$, 700 keV | 28.0 | 0.027 |
| $H^+$, 800 keV | 25.7 | 0.024 |
| $He^+$, 3 keV | 16.6 | 15.5 |
| $He^+$, 30 keV | 53.6 | 4.2 |
| $Ar^{++}$, 60 keV | 500 | 1360 |

$S_n$, nuclear stopping power

$S_e$, electronic stopping power

## 5. Summary

The existence of the 1.65 μm absorption feature in crystalline water ice is an important signature that distinguishes crystalline water ice from amorphous water ice in the outer Solar System. Our experiments showed that the 1.65 μm band of crystalline water ice is weakened by exposure to energetic particles, but not destroyed by it. The survival of the 1.65 μm band is temperature-dependent, showing that there is a balance between the thermal recrystallization and irradiation amorphization. At temperatures of 40 K to 50 K and higher, typical of the middle and outer regions of the known solar system and on Titan's surface, the 1.65 μm band can persist indefinitely despite exposure to space irradiation. This also holds for water ices on the Jovian satellites which hold typical temperatures of 100 K. However, in the Outer Solar system, where ices can be as cold as 10 K (Oort cloud), amorphous water ice likely exists.

## Acknowledgements

This work was financed by the US National Science Foundation (NSF; AST-0507763; DJ, RIK) and by the NASA Astrobiology Institute under Cooperative Agreement NNA04CC08A at the University of Hawaii-Manoa (WZ, DJ, RIK). We are grateful to Ed Kawamura (University of Hawaii at Manoa, Department of Chemistry) for his electrical work.

Figure 1: Infrared reflection-absorption spectra of cubic crystalline water ice recorded at different temperatures between 10 and 140 K. The spectra are offset for clarity.

Figure 2: Infrared reflection-absorption spectra of amorphous water ice recorded at different temperatures between 10 and 50 K. The spectra are offset for clarity.

Figure 3: Infrared reflection-absorption spectra of cubic crystalline water ice before (solid lines) and after (dotted lines) the irradiation at different temperatures. The spectra are offset for clarity.

Figure 4: Recrystallization of the water ice after irradiation. (a) spectrum taken before the irradiation; (b) spectrum taken after 44 hours irradiation; (c) spectrum of the recooled sample after keeping the irradiated sample annealed for one hour at 145K. The spectra are offset for clarity.

Figure 5: The survival of the 1.65 μm band versus the irradiation time at different temperatures.

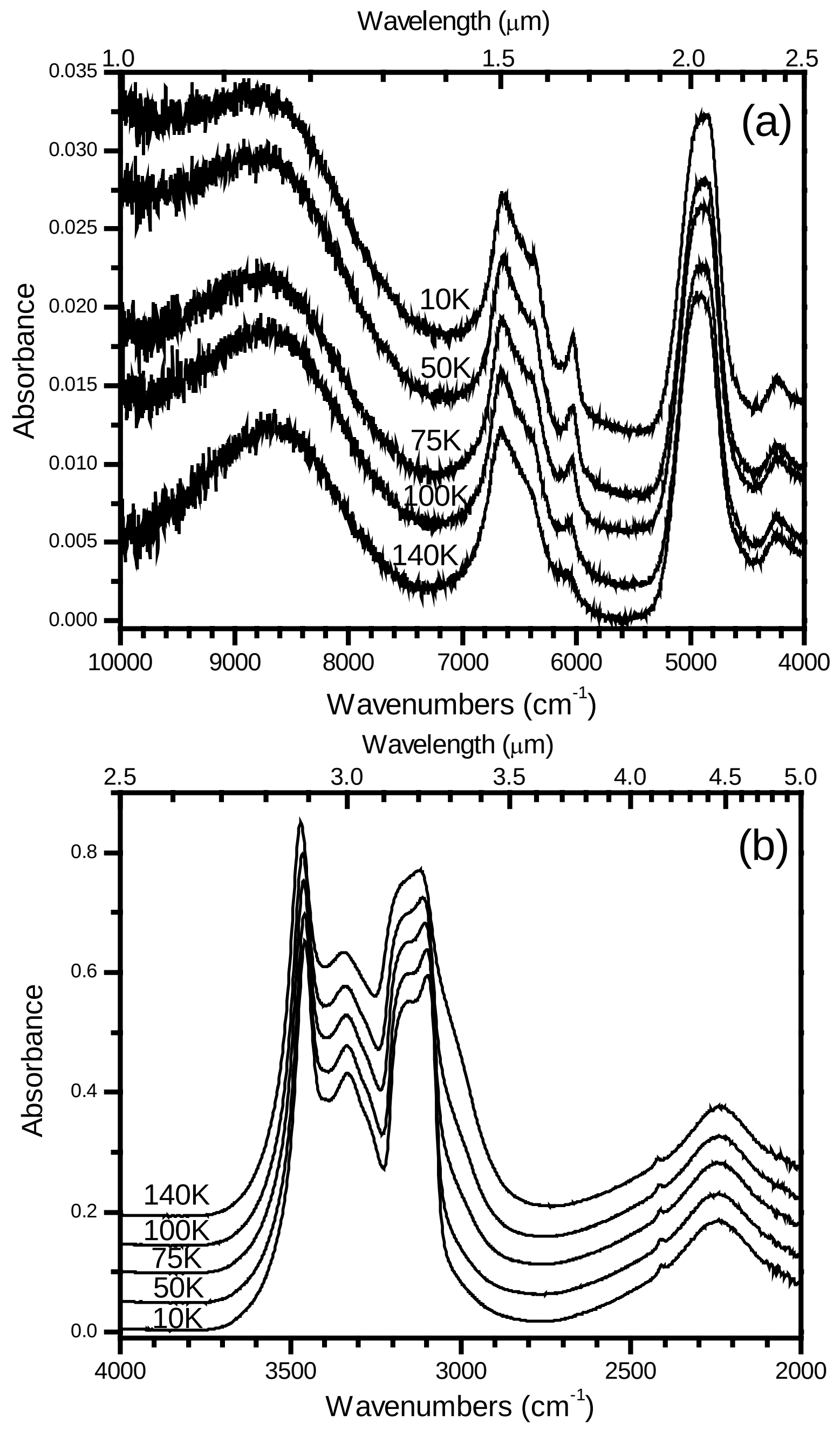

Fig. 1

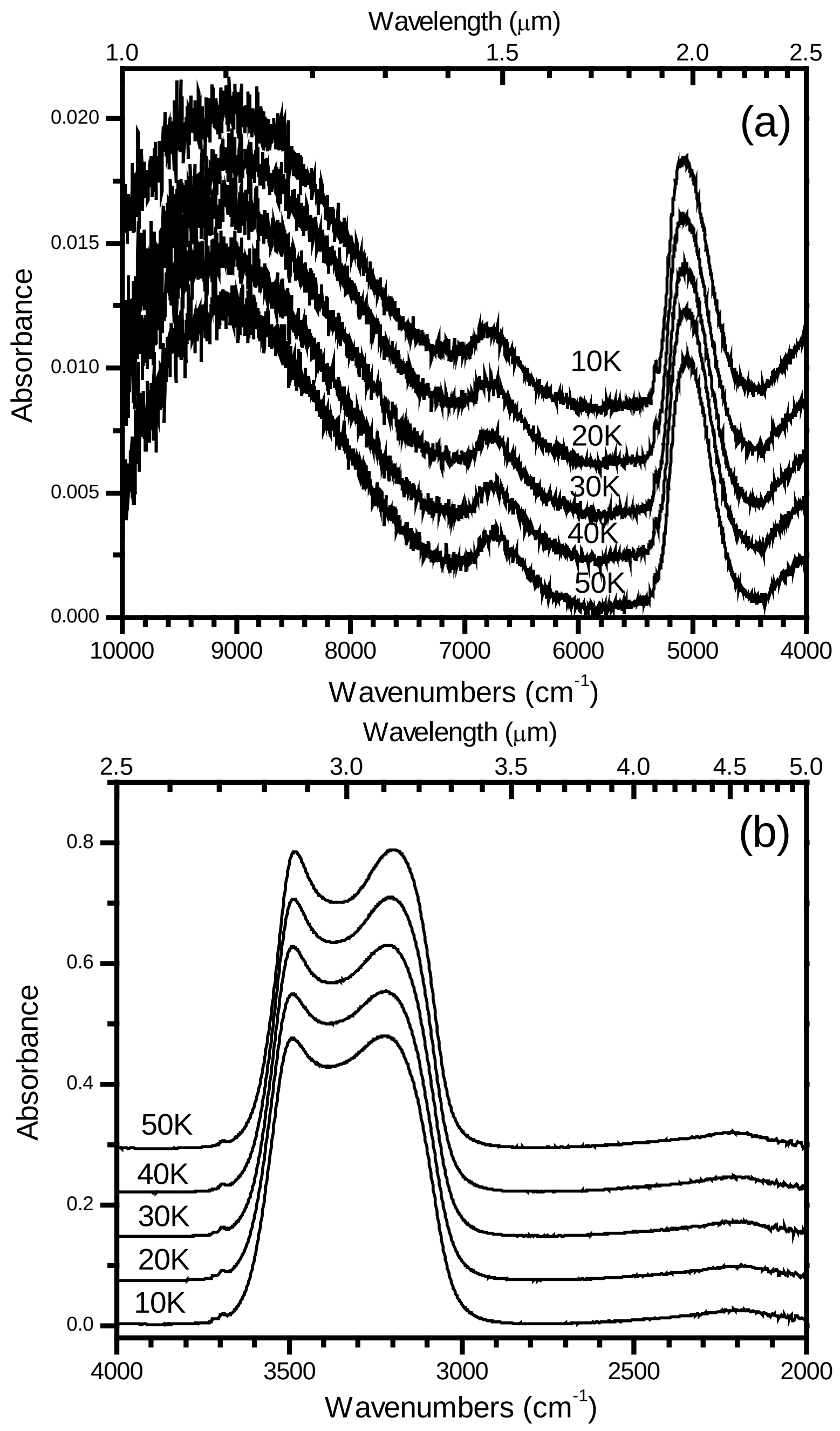

Fig. 2

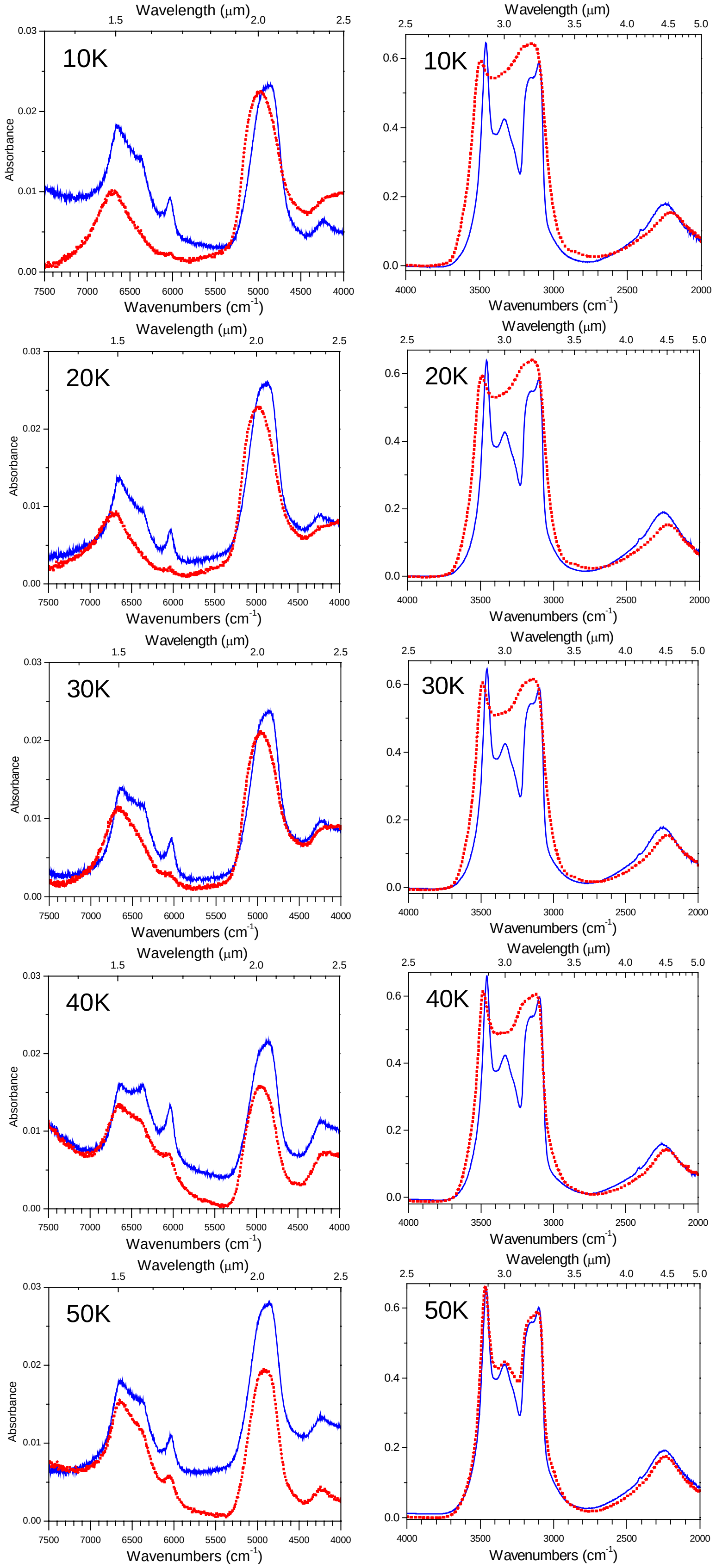

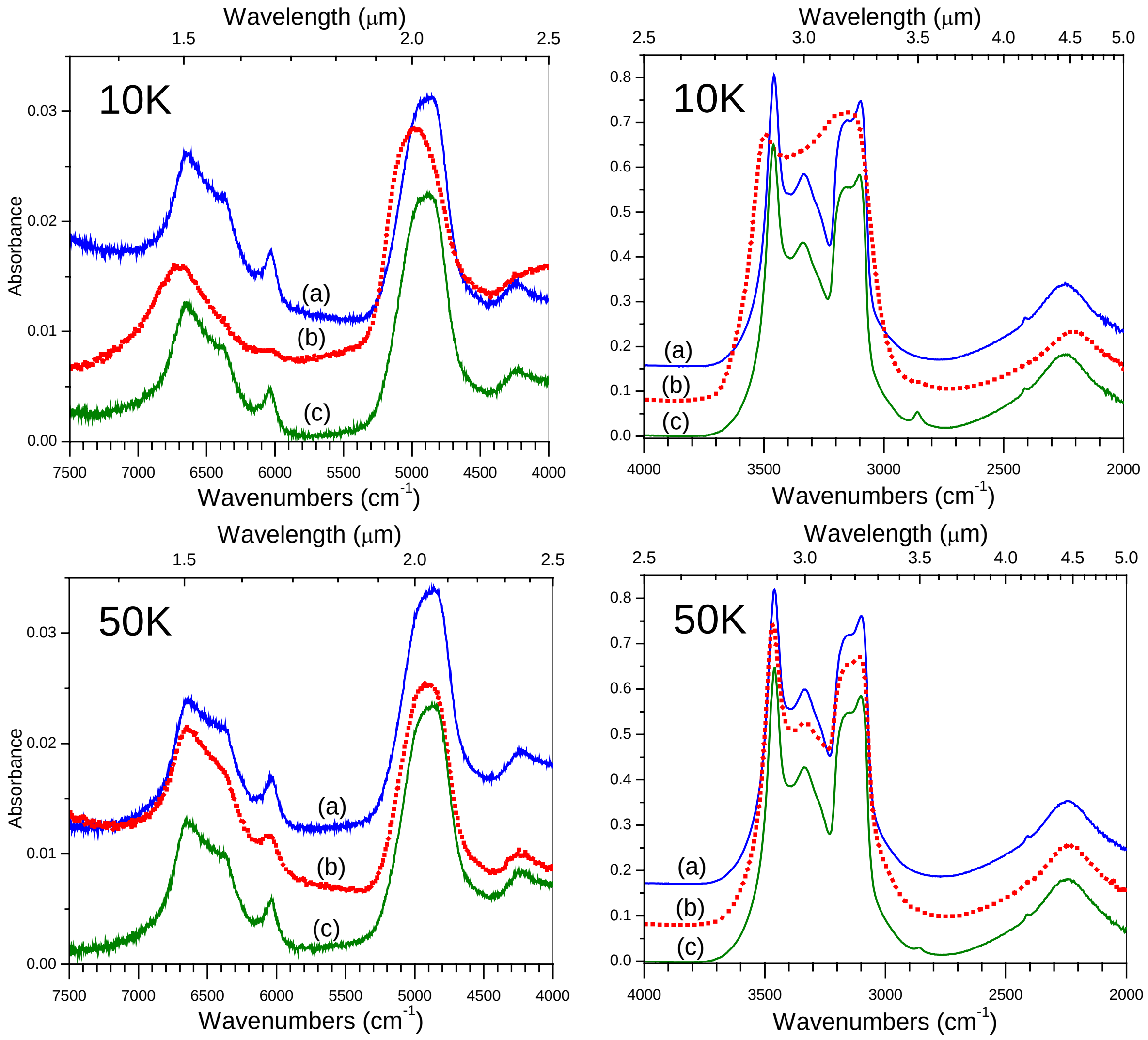

Fig. 4

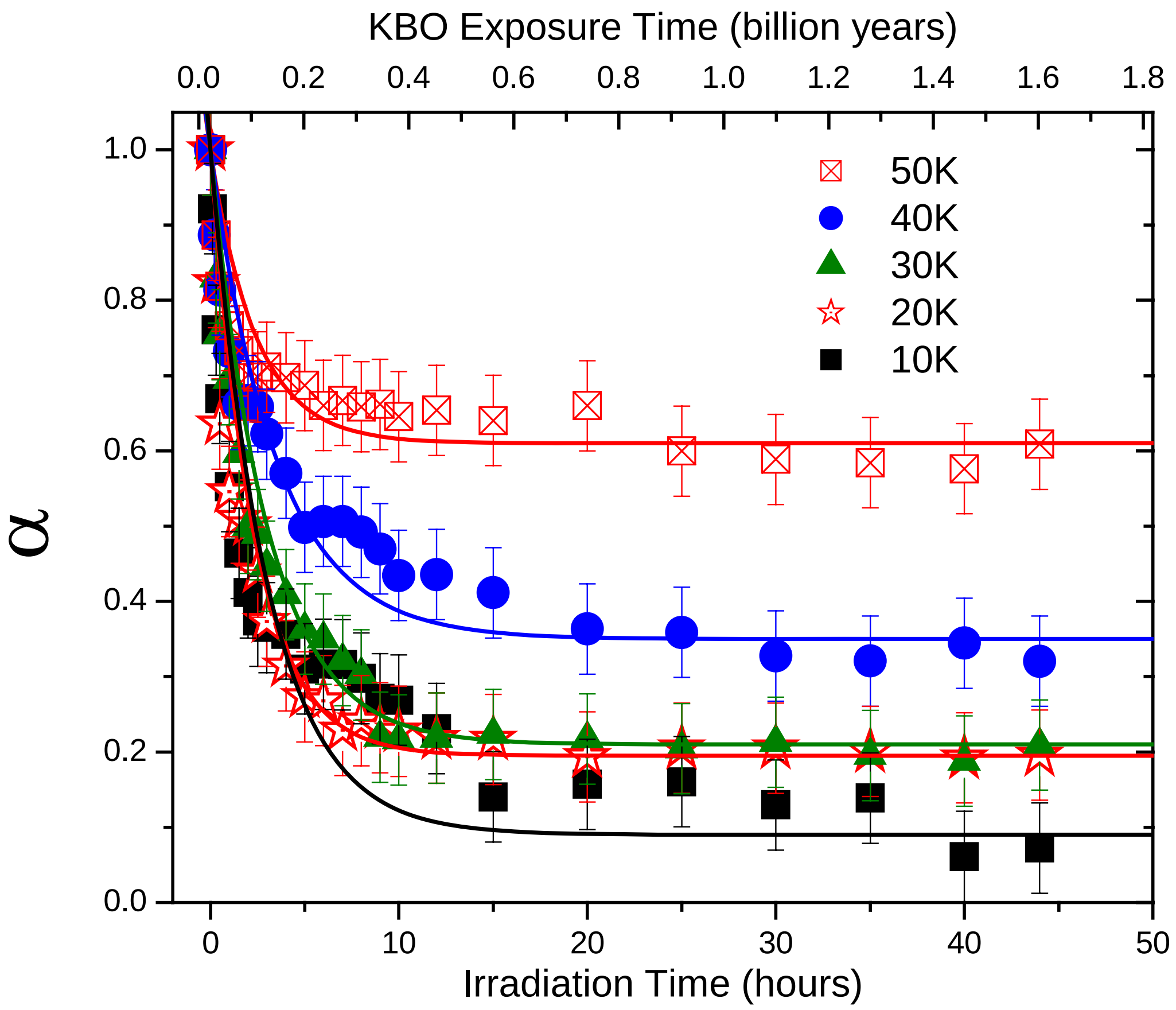

Fig. 5